\documentclass[twocolumn,aps,showpacs]{revtex4}
\usepackage[dvips]{graphicx}

\newcommand{\bra}{\langle}
\newcommand{\ket}{\rangle}
\newcommand{\ketPsi}{| \Psi \rangle}
\newcommand{\braPsi}{\langle \Psi |}
\begin{document}

\title{
% Correlations, fluctuations and entanglement of
% \\ chaotic states in macroscopic quantum systems
Correlations of observables in 
chaotic states of macroscopic quantum systems
}
\author{Ayumu Sugita}
\email{sugita@a-phys.eng.osaka-cu.ac.jp}
\affiliation{
Department of Applied Physics, Osaka City University
3-3-138 Sugimoto, Sumiyoshi-ku, Osaka, 558-8585, Japan}

\author{Akira Shimizu}
\email{shmz@ASone.c.u-tokyo.ac.jp}
\affiliation{
Department of Basic Science, University of Tokyo, 
3-8-1 Komaba, Tokyo 153-8902, Japan and
PRESTO, Japan Science and Technology Corparation, 
4-1-8 Honcho, Kawaguchi, 
Saitama, Japan}

\date{\today}
            
\begin{abstract}
We study correlations of observables 
% chaotic states of macroscopic quantum systems
% , fluctuations and entanglement
in energy eigenstates of 
chaotic systems of a large size $N$.
We show that the bipartite entanglement of two subsystems is 
quite strong, whereas macroscopic entanglement of the total 
system is absent.
% Therefore chaotic states are much more stable than 
% macroscopically entengled states.
% the total system is strongly entangled in some sense, but
% the strong entanglement washes out any 
It is also found that correlations, either quantum 
or classical, among less than $N/2$ points are quite small.
These results imply that chaotic states are stable.
% As a result, fluctuations of any additive operators are
% $O(N)$ or less.
Invariance of these properties under local operations is also shown.

\end{abstract}

\date{\today}

\maketitle

It is generally believed that almost all macroscopic systems 
have chaotic dynamics, 
because otherwise thermodynamics would not hold.
However, properties of chaotic quantum states in 
macroscopic systems are poorly understood
as compared with those in systems of 
small degrees of freedom \cite{qc}.
% which 
% have been studied intensively.
Among such properties, 
correlations of 
observables at different points
are of particular interest.
For example, 
such correlations are directly related to entanglement
\cite{sugita}, 
which is the central subject of quantum information theory
\cite{ent}.
The correlations also define the `cluster property', 
which is one of the most fundamental notions 
of field theory \cite{haag}. %many-body quantum physics.
Moreover, the correlations 
determine the magnitudes of 
fluctuations of `additive operators' (see below),
which are macroscopic observables.
Furthermore, 
Shimizu and Miyadera \cite{SM2002}
(Hereafter referred to as SM)
showed that two-point correlations determine 
the stabilities against classical noises,
perturbations from environments, and local measurements.
% Here, the term `correlation' includes {\em 
% both quantum and classical ones}.
In particular, they showed that 
states with `macroscopic entanglement' are unstable.
Since 
chaotic dynamics is generally believed to promote entanglement
production \cite{Miller}, and since 
classical chaos is characterized by extreme sensitivity to
the initial condition,
it might be tempting
to conjecture that chaotic states would be unstable.
However, as we will show, such a naive expectation is wrong.
% An important point is that there are a huge number of
% possible measures or indices of entanglement for macroscopic states.
% 
% We therefore investigate 
% We show that the bipartite entanglement of two subsystems is 
% quite strong, whereas macroscopic entenglement of the total 
% system is absent.
In this work, we study these points for chaotic quantum
states in macroscopic systems, i.e., systems with a
large but finite degrees of freedom.
% Chaotic dynamics is generally believed to promote entanglement
% production \cite{Miller}.
% Moreover, states with `macroscopic entanglement' are unstable according to SM.
% Furthermore, classical chaos is characterized by extreme sensitivity to
% the initial condition.  
%% Based on these observations, 
% It might thus be tempting
% to conjecture that macroscopic chaotic states would be unstable.
% It is important, however, to note that there are a huge number of
% possible measures or indices of entanglement for macroscopic states.
% We therefore investigate 
% We show that the bipartite entanglement of two subsystems is 
% quite strong, whereas macroscopic entenglement of the total 
% system is absent.
% 
% 
% although chaotic states are strongly (or almost maximally)
% entangled by some measures they are {\it not 
% macroscopically} entangled, and hence are stable, in the sense of SM.

{\em Two-point correlations and fluctuations of additive operators:}
%{\em Correlations in the random matrix theory} 
We consider an energy eigenstate $\ketPsi$ of 
a quantum chaotic system which is composed of $N$ ($\gg 1$) sites,
where 
the Hilbert space is the 
tensor product of local Hilbert spaces.
Let $\{\hat a_{\alpha}(l) \}$ be a basis of local observables at a site $l$.
Assuming that $\hat a_{\alpha}(l)$'s are bounded, we 
normalize them as 
${\rm Tr}[ \hat a_{\alpha}^{\dagger}(l) \hat a_{\beta}(l) ]
= \mbox{constant} \times \delta_{\alpha \beta}$.
Then, all information about the two-point correlations in $\ketPsi$
are included in the 
variance-covariance matrix (VCM),
$ %\begin{eqnarray}
V_{\alpha l, \beta l'} \equiv 
\braPsi 
\delta \hat{a}_{\alpha}^{\dagger}(l)\delta \hat{a}_{\beta}(l')
\ketPsi,
$ %\end{eqnarray}
where $\delta \hat a_{\alpha}(l)\equiv
\hat{a}_{\alpha}(l)-\braPsi \hat{a}_{\alpha}(l) \ketPsi$ \cite{Morimae}.
This matrix also gives us information about 
fluctuations $\braPsi \delta \hat{A}^\dagger \delta \hat{A} \ketPsi$ of 
`additive operators' $\hat A$, which are defined as
the sums of local operators;
$ %\begin{eqnarray}
\hat{A} = \sum_{\alpha,l}c_{\alpha l} \hat{a}_{\alpha}(l).
$ %\end{eqnarray}
Here, $c_{\alpha l}$'s are c-numbers independent of $N$.
Without loss of generality, we here normalize them as
$\sum_{\alpha,l} |c_{\alpha l}|^2 = N$.
As shown in Ref.~\cite{Morimae}, the maximum 
fluctuation of the additive operators is $N e_{\rm max}$,
where $e_{\rm max}$ is the maximum eigenvalue of 
the VCM.
%When $\{ c_{\alpha l} \}$ is taken as the eigenvector 
%$\{c^{(k)}_{\alpha l}\}$ 
%corresponding to the $k$-th eigenvalue $e_{k}$ of the VCM, 
%we find 
%\begin{eqnarray}
% \braPsi \delta\hat{A}_{k}^{\dagger}\delta\hat{A}_{l} \ketPsi 
% = N e_{k}\delta_{kl}.
%\braPsi \delta \hat{A}_{k}^{\dagger} \delta\hat{A}_{k} \ketPsi 
%= N e_{k},
%\end{eqnarray}
%where 
%$%\begin{equation}
%\hat{A}_{k} = \sum_{\alpha,l}c^{(k)}_{\alpha l}\hat{a}_{\alpha}(l).
%$ %\end{equation}
%Therefore, 
%the maximum eigenvalue $e_{\rm max}$ of the VCM gives 
%the maximum value of 
%$\braPsi \delta \hat{A}^{\dagger} \delta \hat{A} \ketPsi$
%as $N e_{\rm max}$, 
%and 
%the eigenvector corresponding to $e_{\rm max}$ gives $\hat A$
%that has this largest fluctuation.
For example, 
$e_{\rm max}=O(1)$ \cite{footnote1}
for any product state 
$\ketPsi = \bigotimes_{l} | \psi_l \ket$, 
whereas 
$e_{\rm max}=o(N)$
for `vacuum states' of many-body physics
in accordance with thermodynamics \cite{SM2002,ruelle}.
Interestingly, 
as special states in large but finite systems, 
there also exist pure states for which $e_{\rm max}=O(N)$ 
\cite{SM2002,Morimae}.
Such pure states are superpositions of macroscopically distinct states,
% have long-range correlations over 
% a region of size $O(N)$, 
hence are called {\em macroscopically entangled states} 
\cite{SM2002,Morimae,Shimizu_Morimae}.
% Such macroscopic entanglement was shown to play a crucial role in speedup of 
% quantum algorithms over classical ones \cite{US2004}.
%\footnote{Note that the fluctuations of additive operators
%are determined by two-point correlations,
%and not by bipartite entanglements.
%For example, a 'cat state' has strong
%two-point correlations and is an AFS,
%though bipartite entanglement between an arbitrary
%pair is zero.}.

Let us evaluate the VCMs of energy eigenstates in chaotic systems using 
random matrix theory (RMT). 
Suppose that an energy 
eigenstate $|\Psi_\lambda \rangle$, labeled by a single parameter $\lambda$, 
is represented as 
$|\Psi_\lambda \rangle = \sum_{i} c_{\lambda i}| i \rangle$
in some basis $\{ |i \rangle \}$.
% $|\lambda\rangle = \sum_{i}c_{\lambda i}|i\rangle$
% in some basis $\{ |i\rangle \}$.
The ensemble averages, denoted by overline, 
of products of the coefficients can be 
calculated easily using RMT \cite{Ullah}. 
For example, in the $d\times d$ GUE or GOE,
\begin{eqnarray}
\overline{|c_{\lambda i}|^{2}} 
& = & \frac{1}{d}, %\;\;\; \left(\frac{1}{d}\right),
\label{fm1}\\
\overline{|c_{\lambda i}|^{4}} & = & 
\frac{2+q}{d(d+1+q)},
\label{fm2}\\
\overline{|c_{\lambda i}|^{2}|c_{\lambda i'}|^{2}} & = & 
\frac{1}{d(d+1+q)}
\ \mbox{ for } i \ne i',
\label{fm3}\end{eqnarray}
where $q = 0$ and $1$ for GUE and GOE, respectively.
Up to the 4-th order, the other combinations which are not written
above vanish identically. 
By using these formulae, we can evaluate 
the averages and variances of the matrix elements of VCMs.
Since the right-hand sides of 
Eqs.~(\ref{fm1})-(\ref{fm3}) are independent of $\lambda$, 
we will hereafter drop the label $\lambda$.
Furthermore, we will denote
$\overline{\bra \Psi_\lambda | \cdot | \Psi_\lambda \ket}$ simply by
$\overline{\bra  \cdot  \ket}$.

\begin{figure}[htbp]
\includegraphics[width=0.45\textwidth]{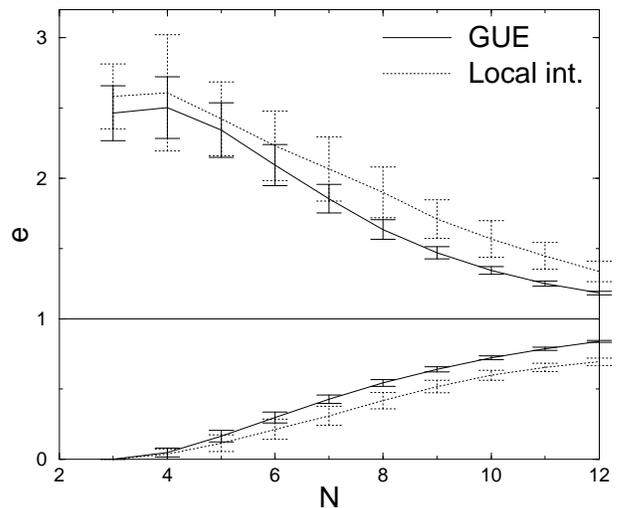}
\caption{The maximum and the minimum eigenvalues of the VCMs in 
spin-$\frac{1}{2}$ systems
as functions of the system size $N$.
The solid lines
represent the results for eigenstates of GUE. 
The results from GOE, which are not displayed here, are very similar to those of GUE.
The dotted lines are the results for the 
the central ($2^{N-1}$-th) eigenstates of the hamiltonian (\ref{Hamiltonian}).
In both cases the averages have been taken over 100 samples.
The error bars show the standard deviation.}
\label{eminmax}
\end{figure}

We here present results for 
the case where each site is a qubit, which is equivalent to 
a spin-$\frac{1}{2}$ system.
In this case, 
we can use the Pauli matrices 
$\{\sigma_{x},\sigma_{y},\sigma_{z}\}$
as a basis of local observables 
\cite{footnote2}.
Therefore, the VCMs of an $N$-qubit system are $3N\times 3N$ matrices.
It is easy to show that 
$
\overline{\langle \sigma_{\alpha}(l) \rangle} = 0
$
and
$
\overline{\langle \sigma_{\alpha}(l) \rangle 
\langle \sigma_{\beta}(l') \rangle} = 0
$ for $l \neq l'$
($\alpha,\beta=x,y,z$). 
With the help of these, 
the average values of the elements of the VCM are calculated as
\begin{eqnarray}
\overline{\langle\delta\sigma_{\alpha}(l)\delta\sigma_{\beta}(l')\rangle}
= \frac{2^{N}}{2^{N}+1+q}\delta_{\alpha \beta}\delta_{ll'}.
\label{corr_spin}
\end{eqnarray}
This suggests that the VCM converges to the unit 
matrix 
$I$
%$\openone$ 
as $N \to \infty$, 
hence $e_{\rm max} = O(1)$.
To examine this point, 
we plot in Fig.~\ref{eminmax} 
the maximum and the minimum eigenvalues
$e_{\rm max}$ and $e_{\rm min}$, of the VCMs
of energy eigenstates of GUE and GOE
as functions of $N$.
It is found that both $e_{\rm max}$ and $e_{\rm min}$
(hence all eigenvalues) concentrate
on $1$ as $N$ increases, and thus 
the VCMs approach 
$I$
%$\openone$
.This means that fluctuations of all additive operators 
are $O(N)$ or less, and that
all two-point correlations for $l \ne l'$ are vanishingly small.
In fact, for $l \ne l'$ we can directly show that 
\begin{eqnarray}
%\overline{
%|\langle\delta\sigma_{\alpha}(l)\delta\sigma_{\beta}(l')\rangle|^{2}}
%\sim
\overline{|\langle\sigma_{\alpha}(l)\sigma_{\beta}(l')\rangle|^{2}}
= \frac{1+q}{2^{N}+1+q},
% = \frac{1}{2^{N}+1}\;\;\;
% \left(\frac{2}{2^{N}+2}\right),
\label{corr_spin2}
\end{eqnarray}
%which %Hence, the correlation between two points 
%decreases exponentially as $N$ increases. 
hence $\overline{|\langle\delta\sigma_{\alpha}(l)\delta
\sigma_{\beta}(l')\rangle|^{2}}$ is also $O(2^{-N})$.
Therefore, 
{\em either entanglement or classical correlation
will be found to be 
exponentially small %infinitesimal
if one observes arbitrary two points of a chaotic state}.
% vanishing two-point correlations.
Thus, in the sense of SM, 
chaotic states have the `cluster property' 
\cite{footnote3},
and are {\em not} entangled macroscopically.
According to the general results of SM, 
these imply that 
chaotic states are stable under local measurements,
and that
the decoherence rate $\Gamma$ of chaotic states is $O(N)$ or less 
under weak perturbations from any noises or environments
which have short time correlations.
This is in sharp contrast to 
macroscopically entangled states 
% (which have $e_{\rm max} =O(N)$) 
for which some noise gives $\Gamma = O(N^2)$ \cite{SM2002}.
% From the general theorem proved by SM, 
% (with a slight generalization as will 
% be discussed later), 
%we conclude also that 
These points will be discussed later again.

{\em Correlations among many points:}
We have seen that two-point correlations are infinitesimal.
% for eigenstates of random matrices.
How about $m$-point correlations for larger $m$?
We can estimate it from bipartite entanglement between two
subsystems as follows.

\begin{figure}[htp]
\includegraphics[width=0.45 \textwidth]{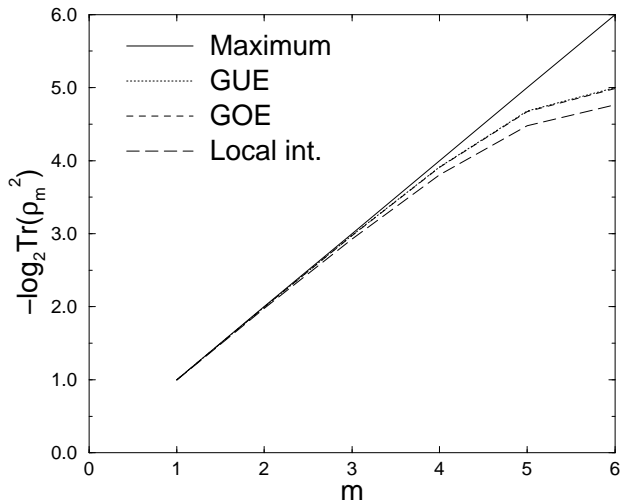}
\caption{
% The logarithm of the average purity
% $\overline{{\rm Tr}\left(\hat{\rho}_{m}^{2}\right)}$
$-\log_{2}\overline{{\rm Tr}\left(\hat{\rho}_{m}^{2}\right)}$
of a subsystem composed of $m$ sites 
is plotted for N=12.
The solid line denotes the upper limit, 
corresponding to the lower limit of 
$\overline{{\rm Tr}\left(\hat{\rho}_{m}^{2}\right)}$, 
 for each $m$.
The dotted and dashed lines, 
which look overlapped each other in this figure, 
represent results for GUE and GOE, respectively.
The long dashed line shows the result obtained 
from the central ($2^{N-1}$-th) eigenvector of the
hamiltonian (\ref{Hamiltonian}). 
}
\label{tr2}
\end{figure}

Let us separate the system into two subsystems A and B
which contain $N_{\rm A}$ and $N_{\rm B} = N-N_{\rm A}$ sites, respectively.
The Hilbert space of subsystem A (B) has dimension $d_{\rm A} =2^{N_{\rm A}}$
($d_{\rm B}=2^{N_{\rm B}}$). We assume that $N_{\rm A}\le N_{\rm B}$, and evaluate `purity'
${\rm Tr}\left(\hat{\rho}_{\rm A}^2\right)$
as a measure of bipartite entanglement,
where $\hat{\rho}_{\rm A}$ is the reduced 
density operator of A;
$\hat{\rho}_{\rm A} \equiv {\rm Tr}_{\rm B} \left( | \Psi \rangle \langle \Psi | \right)$.
\cite{footnote4}
The purity takes the maximum value $1$
when the state $| \Psi \rangle$ 
is separable (unentangled), and the minimum value
$1/d_{\rm A}$ when $\hat{\rho}_{\rm A}$ is a scalar matrix 
%$\openone
$I/d_{\rm A}$.
For GUE and GOE, Eqs.~(\ref{fm1})-(\ref{fm3}) yield the average purity as
\cite{Zyczkowski}
\begin{eqnarray}
\overline{{\rm Tr} \left(\hat{\rho}_{\rm A}^{2}\right)} =
\frac{d_{\rm A}+d_{\rm B}+q}{d_{\rm A}d_{\rm B}+1+q}
% \frac{d_{\rm A}+d_{\rm B}}{d_{\rm A}d_{\rm B}+1}\;\;\; 
% \left(\frac{d_{\rm A}+d_{\rm B}+1}{d_{\rm A}d_{\rm B}+2}\right), 
=\frac{1}{d_{\rm A}} 
\left( 1+
\frac{1}{2^{\Delta N}} + \cdots
% O\left( {1 \over 2^{\Delta N}} \right)
% {1+O(2^{-N_{\rm A}}) \over 2^{\Delta N}}
\right),
% ={1 \over d_{\rm A}} \left(1+2^{-\Delta N}  
% + O \left( {1 \over 2^{N_{\rm B}}} \right) \right)
%={1 \over d_{\rm A}} + O \left( {1 \over d_{\rm B}} \right)
\label{tr2rmt}
\end{eqnarray}
where $\Delta N \equiv N_{\rm B} - N_{\rm A}$ ($\geq 0$), and 
`$\cdots$' denotes smaller terms.
It is found that the average purity is quite small,
which means that {\em the bipartite entanglement of
two subsystems are strong}.
With increasing $\Delta N$, 
in particular, 
the average purity approaches exponentially the minimum value $1/d_{\rm A}$.
Therefore
$\hat{\rho}_{\rm A}$ converges to 
%$\openone
$I/d_{\rm A}$ 
exponentially with increasing $\Delta N$,
for almost all states.
%Therefore, 
%if the total system is divided into 
%two subsystems, they 
%are largely entangled for any sizes 
%$N_{\rm A}$ and $N_{\rm B}$ of them, 
%and maximally entangled when $2^{|N_{\rm B}-N_{\rm A}|} \gg 1$.
This is demonstrated for $N=12$ in Fig.~\ref{tr2}, 
in which $\hat{\rho}_{m} = \hat{\rho}_{\rm A}$ and $m=N_{\rm A}$.
It is noteworthy that the bipartite entanglement is 
large but not maximum when $N_{\rm A}=N_{\rm B}$.

Suppose now that observables 
$\hat{a}_{1}(l_{1}), \hat{a}_{2}(l_{2}), \dots, \hat{a}_{m}(l_{m})$
at $m$ different points (sites) 
$l_1, l_2, \cdots, l_m$ are measured.
Since these observables commute with each other,
their correlations can be calculated in a manner 
similar to the case of classical stochastic variables.
In particular, the cumulants are given by
\begin{eqnarray}
\langle\hat{a}_{1}(l_{1})
\dots\hat{a}_{m}(l_{m})\rangle_{c}
&=&
(-1)^{m}\frac{\partial^{m}\ln Z}
{\partial J_{1}\dots\partial J_{m}},
\label{cummulant}
\end{eqnarray}
where 
$
Z(\{J_{i}\}) \equiv 
\langle \Psi | %\left\langle
\exp\left[-\sum_{i}J_{i}\hat{a}_{i}(l_{i})\right]
| \Psi \rangle %\right\rangle
$
is the generator of moments. 
If we regard the set of $m$ sites $l_1, l_2, \cdots, l_m$ as subsystem A, 
we have 
$
Z(\{J_{i}\})
={\rm Tr} \left( \hat \rho_m
\exp\left[-\sum_{i=1}^m J_{i}\hat{a}_{i}(l_{i})\right] \right)
$,
where $\hat \rho_m = \hat \rho_{\rm A}$.
% is determined by the reduced density operator $\hat \rho_{\rm A}$ of the 
% subsystem.
%
When $4^{m - N/2} \ll 1$, 
% When $2^{2m} \ll 2^{N}$, 
i.e., when $2^m \ll 2^{N-m}$, 
$\rho_m=
%\openone
I/2^m$ to a good approximation, 
according to our results above.
In this case, 
we obtain
$
\langle\hat{a}_{1}(l_{1})
\dots\hat{a}_{m}(l_{m})\rangle_{c}=0
$,
which means that there is no correlation
among $a_{1}(l_{1}),\dots,a_{m}(l_{m})$.
Since they are arbitrary observables at $m$ points, 
it is concluded that 
{\em neither entanglement nor classical
correlation can be found % when $2^{2m - N} \ll 1$.
if $4^{m - N/2} \ll 1$}, i.e., 
if the number $m$ of points one performs measurements
is somewhat smaller than % $N/2$. 
half the total number of points.
% sufficiently smaller than the number of total degrees
% of freedom. 
%

On the other hand, if observables at all points are measured, 
one can detect all types of entanglement.
For example, one can detect
the bipartite entanglement by measuring 
a correlation of the 
Clauser-Horne-Shimony-Holt (CHSH) type \cite{CHSH}.
% takes larger values for entangled states 
% than for separable states or for local classical theories. 
% The correlation 
This correlation becomes almost maximum for chaotic states,
because we have already shown that the bipartite entanglement between
subsystems A and B is almost maximal (when $2^{\Delta N} \gg 1$).
% In this sense, one can say that chaotic states 
% of macroscopic systems have strong entanglement and 
% strong
% many-point correlations. 
However, the enhancement of the CHSH correlation 
over that of separable states (or of local classical theories) 
is
at most by a factor of $\sqrt{2}$.
Recently, 
another correlation of local observables
was found in Ref.~\cite{Shimizu_Morimae},
which detects macroscopic entanglement in general states
(either pure or mixed):
It becomes
as large as $O(N^2)$ for macroscopically entangled states, 
% for which $e_{\rm max}=O(N)$, 
whereas
it is at most $O(N)$ for states with $e_{\rm max}= O(1)$
(and for any mixtures of such states).
For chaotic states, 
this correlation is at most $O(N)$
because we have shown that $e_{\rm max}= O(1)$.
% since . 
% In this sense, 
% entanglement of chaotic states of macroscopic systems is {\em weak}.
%As $m$ is increased to larger than $N/2$, 
% the result of the previous section shows that 
%certain $m$-point correlations grow, until,
%according to our results above,
%certain $N$-point correlations become 
%quite large for $m=N$. 
%Although the result for
%GUE in (\ref{tr2rmt}) has appeared several times
%in the literature \cite{Zyczkowski}, the result
%for GOE is new, and the correlations of local observables
%have evaluated first in this paper by using these formulae.
% Therefore, macroscopic entanglement is absent in 
% chaotic states of macroscopic systems, although 
% the bipartite entanglement is large.
% the strength of entanglement of chaotic states depends 
% crucially on the measures or indices of entanglement.
% The same conlusion was obtained also for other many-body states in 
% Ref.~\cite{Morimae}.

{\em A model with short-range interactions:}
We have assumed RMT in the above analysis.
One might suspect that RMT would not be fully applicable 
to macroscopic systems, 
in which interactions are short-ranged, 
because
RMT assumes that strengths of interactions are 
of the same order between any two points.  
To explore this point, 
we study eigenstates of the following hamiltonian:
\begin{eqnarray}
H &=& 
J \sum_{l=1}^{N} \big\{
\sigma_{x}(l)\sigma_{x}(l+1)
+ \sigma_{z}(l)\sigma_{z}(l+1)
\nonumber \\
&&  \qquad
+ \sqrt{2} \cos \phi_{l} \, \sigma_{y}(l)\sigma_{y}(l+1)
\big\}
\nonumber \\
&&    
- h \sum_{l=1}^{N} \{ \sin\theta_{l} \, \sigma_{x}(l)
+      \cos\theta_{l} \, \sigma_{z}(l) \}.
%
% \sum_{l=1}^{N} \big[
% J \sum_{\alpha =x,z}
% \sigma_{\alpha}(l)\sigma_{\alpha}(l+1)
% \nonumber \\
%  &&    
% + \sqrt{2} J \cos \phi_{l} \sigma_{y}(l)\sigma_{y}(l+1)
% \nonumber \\
% &&    
%       - h \{ \sin\theta_{l} \, \sigma_{x}(l)
%       +      \cos\theta_{l} \, \sigma_{z}(l) \}
% \big].
% J\sum_{l=1}^{N}\{
% \sigma_{x,l}\sigma_{x,l+1}
%       + \sqrt{2}\sigma_{y,l}\sigma_{y,l+1} \cos \phi_{j} 
%       +         \sigma_{z,l}\sigma_{z,l+1}
% \} \nonumber \\
%  &&    
%       - h \sum_{l=1}^{N}\{\sigma_{x,l}\sin\theta_{j} 
%       +                   \sigma_{z,l}\cos\theta_{j}
% \}.
\label{Hamiltonian}
\end{eqnarray}
Here, 
$J$ and $h$ are constants, 
 $\{\phi_{l}\}$ and $\{\theta_{l}\}$ are random variables
with $0\le \phi_{l}, \theta_{l}< 2\pi$, 
and $\sigma_{\alpha}(N+1) =\sigma_{\alpha}(1)$.
This hamiltonian describes a one-dimensional spin system 
in which spins interact 
only with their nearest neighbors, 
where the interaction in the $y$-direction is random.
% and $J$ represents average strength of the interaction.
Besides, there is an external magnetic field with strength 
$h$, whose direction is random in the $x$-$z$ plane.
When $J$ and $h$ are sufficiently large
(say, $J=h=1$),
% the eigenstates of this $H$ 
this system becomes chaotic,
except for states around the lower and upper limits of the 
energy spectrum, 
in the sense that the level spacing distribution 
coincides with that of GOE.

The dotted lines in Fig.~\ref{eminmax} represent the maximum
and the minimum eigenvalues of the VCM for
the $2^{N-1}$-th eigenstate, which is located at the center
of the spectrum, where the chaotic nature appears most clearly. 
% The results have been averaged over 
% 10 samples of random variables $\{\phi_{l}\}$ and $\{\theta_{l}\}$.
It is seen that the results agree fairly with those of RMT.
Furthermore, 
the long dashed lines in Fig.~\ref{tr2} represent
$-\log_{2}\overline{{\rm Tr}\left(\hat{\rho}_{m}^{2}\right)}$
for the same state.
The results are close to those for RMT, especially
when $m$ is small. 
Similar results are obtained also for other eigenstates 
except those near the upper and the lower ends of the spectrum.
Since the density of states has a dominant peak 
at the center of the spectrum,
our conclusions from RMT hold for the great majority of the
eigenstates of the Hamiltonian (\ref{Hamiltonian}).\cite{ends}
We thus see that 
RMT correctly describes correlations in energy eigenstates 
of the chaotic system with short-range interactions.

% In RMT, $\hat{\rho}_A\sim I/d_A$ for any small subsystem A,
% and $I/d_A$ is the canonical distribution with infinite 
% temperature.
%%This means that the state is in high temperature because
%%$I/d_A$ is the canonical distribution with infinite temperature.
% Therefore, when we consider real physical systems, 
% the results from RMT apply only 
% to high temperature
% states, which, in our spin model, locate near the center of the spectrum.
% Nevertheless our conclusions from RMT hold for the great majority of the
% eigenstates of (\ref{Hamiltonian}) because the density of states has a 
% dominant peak
% at the center of the spectrum.
% Systematic analysis of eigenstates near the ends of
% the spectrum will be reported elsewhere.

{\it Invariance under local operations:}
What happens if local operations are 
performed on a chaotic state?
Let 
$\{ | i_l \ket \}$ be a basis of the local Hilbert space at site $l$.
An energy eigenstate $|\Psi^{(N)} \ket$ of 
an $N$-site system can be expanded as
\begin{equation}
|\Psi^{(N)} \ket =
\sum_{i_1} \sum_{i_2} \cdots \sum_{i_N} 
c^{(N)}_{i_1 i_2 \cdots i_N} | i_1 \ket | i_2 \ket \cdots | i_N \ket.
\end{equation}
For GUE, the probability density of the $2^N$ coefficients 
$c^{(N)}_{i_1 i_2 \cdots i_N}$ is given by \cite{Ullah}
\begin{eqnarray}
% p(\{ c^{(N)}_{i_1 \cdots i_N} \})
% =
{(2^N-1)! % \Gamma (2^N) 
\over \pi^{2^N}}
\delta\left(\sum_{i_1} \cdots \sum_{i_N} 
|c^{(N)}_{i_1 \cdots i_N} |^{2}-1\right).
%\Gamma (2^N)/\pi^{2^N}
\label{pGUE}\end{eqnarray}
Since this density is invariant under changes of the local basis
$\{ | i_l \ket \}$ for any $l$, the statistical properties 
of $|\Psi^{(N)} \ket$ are not changed by local unitary 
transformations.
% On the other hand, 
Furthermore, 
when a local projective measurement that diagonalizes 
$\{ | i_N \ket \}$ is performed on the $N$-th qubit, 
the post-measurement state (for each measurement) is
given by 
$ %\begin{equation}
| i_N \ket
\sum_{i_1} \cdots \sum_{i_{N'}} 
c^{(N')}_{i_1 \cdots i_{N'}}  | i_1 \ket \cdots | i_{N'} \ket
\equiv
| i_N \ket
|\Psi^{(N')} \ket,
$ %\end{equation}
where 
$N' \equiv N-1$, and
$
c^{(N')}_{i_1 \cdots i_{N'}}
\equiv
c^{(N)}_{i_1 \cdots i_{N'} i_N}
/(\sum_{i_1} \cdots \sum_{i_{N'}} 
|c^{(N)}_{i_1 \cdots i_{N'} i_N} |^{2})^{1/2}
$.
It is easy to show that 
$\{ c^{(N')}_{i_1 \cdots i_{N'}} \}$ also obeys the probability 
distribution of GUE, i.e., 
Eq.~(\ref{pGUE}) with $N$ being replaced with $N'$.
Therefore, all the results for 
$|\Psi^{(N)} \ket$ hold  
for $|\Psi^{(N')} \ket$ as well, with $N$ being replaced with $N'$ 
\cite{footnote5}.
The same can be said 
when a local projective measurement that diagonalizes 
another local basis $\{ | j_N \ket \}$ is performed on the $N$-th qubit, 
because the density (\ref{pGUE}) is invariant under changes of the local 
basis.
The same conclusions can be derived for GOE as well.
We therefore conclude that 
{\em properties of chaotic states which we have found in this work
are invariant under `local operations,'}
including local unitary transformations
and local projective measurements diagonalizing a local basis.
This implies, for example, that a chaotic 
state cannot be disentangled by
% a small number ($\ll N$) of 
local measurements at less than $N$ points.
This is in sharp contrast to, e.g., a `cat state,'
$
|{\rm C} \ket
\equiv
|\uparrow \uparrow \cdots \uparrow \ket / \sqrt{2}
+
|\downarrow \downarrow \cdots \downarrow \ket/ \sqrt{2}
$,
for which a single local measurement suffices to disentangle 
the state \cite{SM2002}.
This is consistent with the conclusion, which we 
have derived above using the theorem by SM, that 
chaotic states are stable under local measurements
because two-point correlations are infinitesimal.

% {\it Degree of entanglement:}
{\it Discussions:}
If one defines the `degree' (or `strength') ${\cal E}$ of 
entanglement % a state 
by the minimum number of local operations 
that are necessary to disentangle it, then,
according to our results, ${\cal E}$ of 
chaotic states in large systems is quite high.
However, this does {\em not} mean that they are anomalous 
as macroscopic states.
Indeed, we have shown that 
they are stable against local measurements and 
that fluctuations of all additive operators are 
$O(N)$ or less,
in accordance with thermodynamics \cite{SM2002,Morimae,ruelle}.
% which is smaller than the upper limit $o(N^2)$ 
% that is assumed in thermodynamics  for pure phases.
%
Furthermore, % according to our results,
as we have mentioned above, 
the correlation of Ref.~\cite{Shimizu_Morimae}, which 
detects macroscopic entanglement, is quite small for chaotic states.
Moreover, in the infinite-size limit, 
all multi-point correlations vanish 
since the number $m$ of measured points is always finite
(hence $m \ll N/2$ as $N \to \infty$).
% Here, the term `correlation' includes both quantum and classical ones.
%Note that the correlations we have calculated 
%have both quantum and classical natures.
Although the absence of {\em quantum} correlations for finite $m$ 
at $N \to \infty$ has been 
generally proved \cite{Shimizu_Morimae}, 
we have found here that chaotic states do not even have 
{\em classical} correlations in the same limit.
Therefore, 
% either entanglement or classical correlation 
correlations, either quantum or classical, of chaotic states are 
neither detectable nor usable in infinite systems.

% In this sense, 
% the entanglement of chaotic states is small, 
% although ${\cal E}$ of the above sense is quite large.
% This two-sided nature is characteristic to 
% entanglement of 
% chaotic states in large systems.

{\it $N$-dependences:}
% In the above analyses, 
In this work, 
we have often discussed $N$ dependences
of various quantities.
Unlike a uniform state in a uniform system, however, 
correspondence between states in systems of different sizes
is non-trivial in chaotic systems.
For completeness, 
we finally describe what the $N$ dependences mean in this paper.

For each system size $N$, consider a {\em set} 
S of all
energy eigenstates whose energies are in an interval
$(E-\Delta/2, E+\Delta/2]$.
To see the $N$ dependence of some quantity $Q$, 
look at its (probability) distribution $P_N(Q)$
in S. %$S[E-\Delta/2, E+\Delta/2)$.
Then, the $N$ dependence of $Q$ can be discussed 
in terms of the $N$ dependence of $P_N(Q)$.
For example, 
we say `$Q \leq Q_0$ for almost all states in S for sufficiently
large $N$' 
if for arbitrary positive number $\epsilon$ there exists
$N_0$ such that 
$%\begin{equation}
1 - P_N(Q \leq Q_0) \leq \epsilon
% \quad \mbox{for $N \geq N_0$}.
$ %\end{equation}
for $N \geq N_0$.
We can thus discuss 
$N$ dependences of various quantities for 
chaotic states.
In particular, we can apply the results of SM, 
in which $N$ dependences play crucial roles.
Note that 
$P_N(Q)$ is independent of $E$ and $\Delta$
in the case of RMT.
% and we can discuss the $N$ dependence without 
% specifying $E$ and $\Delta$.
In the short-range interaction model given by 
Eq.~(\ref{Hamiltonian}), 
% on the other hand, 
the dependence of $P_N(Q)$ on $E$ and $\Delta$
is quite weak if we confine ourselves 
to states around the center of the energy spectrum.
In either case, 
we can discuss the $N$ dependence without 
specifying $E$ and $\Delta$.

% In summary, 
% for energy eigenstates of chaotic systems of a large size $N$, 
% the total system is strongly entangled in some senses. 
% However the strong entanglement washes out 
% any correlations among less than $N/2$ points, 
% and correlations can be observed only when 
% more than $N/2$ points are observed simultaneously.
% As a result, fluctuations of any additive operators
% are O(N) or less.
% These properties, which are expected from RMT, 
% have been confirmed in a model with local interactions.
% These results imply, for example, that 
% chaotic states are much more stable 
% against external perturbations, such as noises
% and local measurements, than
% anomalous macroscopic states such as `cat states.'
% It will be interesting to confirm 
% the present results by experiments.

 Acknowledgment
We thank H. Matsuful for discussions.
A. Sugita was supported by Japan Society for the Promotion of Science
for Young Scientists.


\begin{thebibliography}{99}
\bibitem{footnote1}
Throughout this paper, we say that $f(N)=O(g(N))$
if $\lim_{N\rightarrow\infty} f(N)/g(N) = {\rm const}\ne 0$,
and that $f(N) = o(g(N))$ if 
$\lim_{N\rightarrow\infty} f(N)/g(N) = 0$.
\bibitem{footnote2}
Actually we need four matrices 
$\sigma_{x},\sigma_{y},\sigma_{z}$ and 
%$\openone$
$I$
to represent all $2\times 2$ % Hermitian 
matrices as their linear combinations.
However, we don't have to consider 
%$\openone$
$I$
because all correlations including it vanish.
\bibitem{footnote3}
The cluster property of SM is defined by  
two-point correlations, whereas the one of Ref.~\cite{haag}
is defined by multi-point correlations.
Clearly, a state having the latter cluster property 
has the former one as well.
\bibitem{footnote4}
For many-body states, there are many sorts of entanglement
and, accordingly, many measures or indices of entanglement.
The simplest entanglement is bipartite entanglement, 
in which a many-body system is regarded as a two-body system
composed of two subsystems of the many-body system.
For pure states, the von Neumann entropy 
of the reduced density operator is the unique measure of 
the bipartite entanglement, whereas 
the purity is a simplified version of it.
\bibitem{ends}
Systematic analysis of eigenstates near the ends of
the spectrum will be reported elsewhere.
\bibitem{footnote5}
$|\Psi^{(N)} \ket$ is an energy eigenstate of the $N$-site system, whereas 
$|\Psi^{(N')} \ket$ is not necessarily an energy eigenstate of 
the $N'$-site system.
This difference does not make any difference in 
the statistical properties.
\end{thebibliography}
\end{document}